\documentstyle[12pt]{article}

\begin{document}

\begin{titlepage}
\begin{flushright} Preprint KUL-TF-93/37 \\
                   hep-th/9309084\\
                   September 1993 \\
\end{flushright}

\vfill
\begin{center}
{\large\bf Improved Collective Field Formalism for\\
          an Antifield Scheme for \\
          Extended BRST Symmetry   }\\
\vskip 27.mm
{\bf Frank De Jonghe$^{1,2}$ }\\
\vskip 1cm
Instituut voor Theoretische Fysica
        \\Katholieke Universiteit Leuven
        \\Celestijnenlaan 200D
        \\B-3001 Leuven, Belgium\\[0.3cm]
\end{center}
\vfill
\begin{center}
{\bf Abstract}
\end{center}
\begin{quote}
\small
We present a new collective field formalism with two rather than one
collective field to derive the antifield formalism for extended BRST
invariant quantisation. This gives a direct and physical proof of the
scheme of Batalin, Lavrov and Tyutin, derived on algebraic grounds.
The importance of the collective field in the quantisation of open
algebras in both the BRST and extended BRST invariant way is stressed.
\end{quote}
\vspace{2mm}
\vfill
\hrule width 3.cm  {\footnotesize
\noindent $^1$ Aspirant N.F.W.O., Belgium\\
\noindent $^2$ e-mail: frank\%tf\%fys@cc3.kuleuven.ac.be}
\normalsize
\end{titlepage}

\parindent 5truemm
\parskip 0truemm

\newcommand{\dirac}[1]{/ \!\!\!#1}
\newcommand{\vgl}[1]{eq.(\ref{#1})}
\newcommand{\gv}{\gamma^5}
\newcommand{\gu}[1]{\gamma^{#1}}
\newcommand{\gd}[1]{\gamma_{#1}}

\newsavebox{\uuunit}
\sbox{\uuunit}
    {\setlength{\unitlength}{0.825em}
     \begin{picture}(0.6,0.7)
        \thinlines
        \put(0,0){\line(1,0){0.5}}
        \put(0.15,0){\line(0,1){0.7}}
        \put(0.35,0){\line(0,1){0.8}}
       \multiput(0.3,0.8)(-0.04,-0.02){12}{\rule{0.5pt}{0.5pt}}
     \end {picture}}
\newcommand {\unity}{\mathord{\!\usebox{\uuunit}}}
\newcommand{\half}{{\textstyle\frac{1}{2}}}
%
\newcommand{\dr}{\raise.3ex\hbox{$\stackrel{\leftarrow}{\delta }$}}
\newcommand{\dl}{\raise.3ex\hbox{$\stackrel{\rightarrow}{\delta}$}}
%
\newcommand{\cA}{{\cal A}}
\newcommand{\cB}{{\cal B}}
\newcommand{\cC}{{\cal C}}
\newcommand{\cD}{{\cal D}}
\newcommand{\cE}{{\cal E}}
\newcommand{\cF}{{\cal F}}
\newcommand{\cG}{{\cal G}}
\newcommand{\cH}{{\cal H}}
\newcommand{\cI}{{\cal I}}
\newcommand{\cJ}{{\cal J}}
\newcommand{\cK}{{\cal K}}
\newcommand{\cL}{{\cal L}}
\newcommand{\cM}{{\cal M}}
\newcommand{\cN}{{\cal N}}
\newcommand{\cO}{{\cal O}}
\newcommand{\cP}{{\cal P}}
\newcommand{\cQ}{{\cal Q}}
\newcommand{\cR}{{\cal R}}
\newcommand{\cS}{{\cal S}}
\newcommand{\cT}{{\cal T}}
\newcommand{\cU}{{\cal U}}
\newcommand{\cV}{{\cal V}}
\newcommand{\cW}{{\cal W}}
\newcommand{\cX}{{\cal X}}
\newcommand{\cY}{{\cal Y}}
\newcommand{\cZ}{{\cal Z}}
%
\newcommand{\bA}{{\bf A}}
\newcommand{\bB}{{\bf B}}
\newcommand{\bC}{{\bf C}}
\newcommand{\bD}{{\bf D}}
\newcommand{\bE}{{\bf E}}
\newcommand{\bF}{{\bf F}}
\newcommand{\bG}{{\bf G}}
\newcommand{\bH}{{\bf H}}
\newcommand{\bI}{{\bf I}}
\newcommand{\bJ}{{\bf J}}
\newcommand{\bK}{{\bf K}}
\newcommand{\bL}{{\bf L}}
\newcommand{\bM}{{\bf M}}
\newcommand{\bN}{{\bf N}}
\newcommand{\bO}{{\bf O}}
\newcommand{\bP}{{\bf P}}
\newcommand{\bQ}{{\bf Q}}
\newcommand{\bR}{{\bf R}}
\newcommand{\bS}{{\bf S}}
\newcommand{\bT}{{\bf T}}
\newcommand{\bU}{{\bf U}}
\newcommand{\bV}{{\bf V}}
\newcommand{\bW}{{\bf W}}
\newcommand{\bX}{{\bf X}}
\newcommand{\bY}{{\bf Y}}
\newcommand{\bZ}{{\bf Z}}
\newcommand{\beq}{\begin{equation}}
\newcommand{\eeq}{\end{equation}}
\newcommand{\gras}[1]{\epsilon_{#1}}
\newcommand{\gh}[1]{\mbox{gh} \left( #1 \right)}
\newcommand{\sdet}{\mbox{sdet}}
\newcommand{\str}{\mbox{str}}
\newcommand{\tr}{\mbox{tr}}
\newcommand{\ihbar}{\frac{i}{\hbar}}
\def\sepand{\rule{14cm}{0pt}\and}
\def\gtwid{\raise.3ex\hbox{$>$\kern-.75em\lower1ex\hbox{$\sim$}}}
\def\ltwid{\raise.3ex\hbox{$<$\kern-.75em\lower1ex\hbox{$\sim$}}}

\textwidth 170mm
\textheight 230mm

\topmargin -0.8cm

\section{Introduction}

Today, the best Lagrangian quantisation scheme that is capable of
quantising
any gauge theory of all known types, is the antifield formalism or the
Batalin-Vilkovisky \cite{BV} quantisation scheme. Let us give a non
exhaustive list of assets of this scheme. The antifields that are
introduced, are sources for the BRST transformations. Hence, the
renormalisation of gauge theories as described by Zinn-Justin \cite{Bible}
is naturally incorporated in the BV scheme. Secondly, the antifields allow
you, at least formally, to fix a gauge, do the calculation you are
interested in, and still be able to transform the result to other gauges.
This holds especially for the quantum counterterms needed for maintaining
the Ward identities. The antifields thus prevent you from accidentally
finding a vanishing anomaly (see \cite{ik3} and references therein).
A third, and major, asset of the BV scheme is that the fields and their
associated antifields allow for an algebraic structure which is very
similar to classical mechanics. There is an analog of the Poisson bracket,
the so-called antibracket. The fields and antifields are canonically
conjugated with respect to this bracket. Also, canonical transformations,
i.e. transformations of the fields and the antifields which leave the
antibracket invariant, play
a key part. Gauge fixing can be understood as a canonical transformation
\cite{Siegel}.

Despite all these qualities, the scheme is less used than could be
expected, probably because the meaning of the scheme is somewhat obscured
by its usual algebraic derivation. Recently, Alfaro and Damgaard
\cite{PoulBV} gave an explicit derivation of the BV scheme for closed
algebras starting from ordinary BRST quantisation. This was generalised to
open algebras in \cite{ik2}. The guiding principle \cite{PoulBV} is that
whatever the original BRST algebra, it has to be extended to include the
Schwinger-Dyson BRST symmetry \cite{PoulSD} (below denoted SD-symmetry).
This is implemented using a
collective field, which leads to extra shift symmetries. The antighosts of
this shift symmetry are the BV antifields. This point of view does not only
give more insight in the fact that the BV antifield formalism incorporates
the Schwinger-Dyson equations, it also provides us with an explicit and
intuitive road from ordinary BRST quantisation to the BV scheme.

The same basic idea of extending the BRST algebra with the SD-symmetry has
also been implemented in the Hamiltonian quantisation formalism of Batalin,
Fradkin and Vilkovisky \cite{BFV} in \cite{ik4}. This way, it is possible
to prove the equivalence of the Hamiltonian and Lagrangian formalism. Let
us note in passing that, contrary to what is stated in \cite{ik4}, the
proof is valid for open algebras.

Although BRST--anti-BRST (or extended BRST) quantisation \cite{antiBRST}
allows apparently for nothing more than ordinary BRST quantisation, there
has been a growing interest in it. In \cite{BLT}, Batalin, Lavrov and
Tyutin developed an antifield formalism for extended BRST invariant
quantisation on algebraic grounds. This was partly motivated by the hope to
construct a superfield formulation for the quantisation of general gauge
theories \cite{Lavrov}. In this antifield scheme, one has three rather
than one antifield: a source for the BRST transformation, a source for the
anti-BRST transformation and finally a sourceterm for mixed
transformations. In \cite{ik1}, the collective field approach is used to
derive this antifield formalism. The only difference is the way the
antifields are removed after gauge fixing. In \cite{BLT}, this is done by
fixing them to zero, while in \cite{ik1} a Gaussian integral was used for
that purpose. This latter fact however restricts the validity of \cite{ik1}
to closed algebras \cite{ik2}. In this paper, we show how modifying the
collective field scheme, by introducing two rather then one collective
field, leads to the same gauge fixing procedure as the algebraic approach.
We argue that then indeed open algebras can be quantised and point out the
crucial importance of the collective fields for this.

The plan of this paper is the following. In section 2, we give a short
review of the SD approach to the ordinary BV scheme, in order to make the
analogies and differences with our treatment of the extended BRST case more
clear. In section 3, we introduce our modified collective field technique
and show how it leads to the Schwinger-Dyson equations as Ward identities.
The next section contains the heart of the paper. There we derive the
antifield formalism of \cite{BLT} for closed algebras. Ward identities and
the quantum master equation are discussed in section 5. A comment on open
algebras is given in section 6. Finally, we draw our conclusions.

\section{Collective fields for BV}

In this section we give a very short summary of the collective field
approach for the construction of the BV antifield formalism \cite{PoulBV}.
We start from a classical action $S_0[\phi^i]$, depending on a set of
fields $\phi ^i$. Suppose that this classical action has gauge invariances
which are irreducible and form a closed algebra. Then one can construct a
nilpotent BRST operator acting on an extended set of fields $\phi^A$.
The $\phi ^A$ include the original gauge fields $\phi ^i$,
the ghosts $c^{\alpha}$ and the pairs of trivial systems needed for the
construction of the gauge fermion and for the gauge fixing. We summarise
all their BRST transformation rules by $ \delta \phi^A = \cR^A[\phi^A]$.
We enlarge the set of fields by replacing the field
$\phi^A$ wherever it occurs, by $\phi^A-\varphi^A$. $\varphi^A$ is
called the {\it collective field}. This leads
to a new symmetry, the shift symmetry, for which we introduce a ghost field
$c^A$, and a trivial pair consisting of an antighost field $\phi^*_A$ and
an auxiliary field $B_A$. The BRST transformation rules are given
by:
\begin{eqnarray}
  \label{extBRST}
      \delta \phi^A & = & c^A \nonumber \\
      \delta \varphi^A & = & c^A - \cR^A[\phi -\varphi] \nonumber \\
      \delta c^A & = & 0 \\
      \delta \phi^*_A & = & B_A \nonumber \\
      \delta B_A & = & 0. \nonumber
\end{eqnarray}
Now there are two gauge symmetries to fix. We do this as follows:
\begin{eqnarray}
    \label{SGF}
    S_{gf} & = & S_0[\phi^i-\varphi^i] - \delta [\phi^*_A \varphi^A]
                 + \delta \Psi[\phi ^A] \nonumber \\
           & = & S_0[\phi^i-\varphi^i] + \phi^*_A \cR^A[\phi -\varphi]
           - \phi^*_A c^A + \frac{\dr \Psi}{\delta \phi^A} c^A
           -\varphi^A B_A.  \\
           & = & S_{BV}(\phi -\varphi, \phi^*)
           - \phi^*_A c^A + \frac{\dr \Psi}{\delta \phi^A} c^A
           -\varphi^A B_A.  \nonumber
\end{eqnarray}
This gives the following form of the partition function, which is
typical for the BV scheme:
\beq
   \cZ = \int [d\phi^A] [d\phi^*_A] \delta \left(\phi ^*_A -
      \frac{\dr \Psi[\phi]}{\delta \phi^A} \right)
      e^{\ihbar S_{BV}(\phi, \phi^*)}.
\eeq
The fact that the gauge fixed action is still BRST invariant, leads to the
classical master equation for $S_{BV}$, using that $\delta \varphi^A = c^A
- \frac{\dl S_{BV}(\phi -\varphi )}{\delta \phi^*_A}$ :
\beq
 \frac{\dr S_{BV}(\phi,\phi ^*)}{\delta \phi^A}
       \frac{\dl S_{BV}(\phi, \phi ^*)}{\delta \phi^*_A} = 0.
 \label{ClaMas}
\eeq

Using a BRST invariant action as weight in the partition function, we have
Ward's identities $\langle \delta X \rangle = 0$, for any $X$.
Remember that quantum counterterms may be needed in order to guarantee the
validity of the Ward identities. Imposing that the Schwinger-Dyson
equations should be derivable as Ward identities, restricts their form to
$\hbar M(\phi -\varphi,\phi^*)$. Hence, we replace $S_{BV}(\phi
-\varphi,\phi^*)$ as weight in the path integral by $W(\phi -\varphi ,
\phi^*) = S_{BV}(\phi -\varphi,\phi^*) +\hbar M(\phi -\varphi,\phi^*)$.
Considering quantities $X(\phi ,\phi^*)$, and integrating out all fields of the
collective field formalism, except these, this gives the Ward identity
\beq
   0  =  \int [d\phi ][d\phi^*] \left[ (X,W) - i \hbar \Delta X \right]
          e^{\ihbar W(\phi ,\phi^*)} \delta \left(\phi ^*_A -
      \Psi_A \right),
\eeq
with the antibracket defined by
\beq
   (F , G) = \frac{\dr F}{\delta \phi^A} \frac{\dl G}{\delta \phi^*_A} -
 \frac{\dr F}{\delta \phi^*_A} \frac{\dl G}{\delta \phi^A},
 \label{antibracket}
\eeq
and with the delta-operator
\beq
   \Delta X = (-1)^{\gras{A}+1} \frac{\dr}{\delta \phi^*_A}
\frac{\dr}{\delta \phi^A} X  = (-1)^{\gras{X}}
   (-1)^{\gras{A}} \frac{\dl}{\delta \phi^*_A}
\frac{\dl}{\delta \phi^A} X                   .
\eeq
We also denoted $\frac{\dr \Psi}{\delta \phi^A} = \Psi_A$.

Removing all derivatives from $X$, the Ward identity can be used to derive
the quantum master equation, as it should hold for any $X$.
We get
\beq
   i\hbar \int [d\phi][d\phi^*] X(\phi,\phi^*_A +\Psi_A) \Delta \exp
      \left[ \ihbar W(\phi ,\phi^*_A +\Psi_A) \right] \delta(\phi^*) = 0.
\eeq
and thus the quantum master equation
\beq
   \Delta \exp \left[ \ihbar W(\phi ,\phi^*_A +\Psi_A) \right]  = 0.
\eeq

Let us end this far to short overview with a comment on open algebras. In
\cite{Open}, De Wit and van Holten gave a recipe to construct a BRST
invariant action for gauge symmetries with an open algebra. It consists
in modifying the BRST transformation rules and the action itself by
adding an expansion to both in powers of the derivatives of the gauge
fermion with respect to the gauge fields. For the case of the gauge fermion
in the collective field
formalism, $F = \phi^*_A \varphi^A +\Psi[\phi]$, we thus have to expand
in powers of $\phi^*_A$ and $\Psi_A$. A solution
which is linear in the latter can be found, and only an expansion in the
antifield remains. This way, the form of $S_{BV}$ for open algebras, that
is, an extended action with terms that are of quadratic or higher order in
the antifields, is recovered. For more details, see \cite{ik2}.

A posteriori, the collective field formalism can be seen as a justification
\footnote{This point of view was stressed by P.H. Damgaard.} of the
procedure of De Wit and van Holten. We will develop this point of view here
in some detail, as it will be our starting point for the treatment of open
algebras in the extended BRST antifield formalism. When quantising a
gauge theory, one always has to choose a set of functions $F^{\alpha}$,
defining a gauge. This is at least so for every known scheme today. The
quantisation should at least satisfy the following three requirements.
(i) The partition function and expectation values should be well-defined,
owing to a careful choice of the functions $F^{\alpha}$. This is the {\it
admissability condition} for the gauge fixing. (ii) Although defined using
specific $F^{\alpha}$, the partition function  should be invariant under
(infinitesimal) deformations of the functions $F^{\alpha}$, i.e. gauge
independent. (iii) When putting the $F^{\alpha}$ to zero in the
expressions for the partition function and expectation values, they should
reduce to the ill-defined expressions one started from.

The introduction of collective fields allows us to construct the BRST
transformation rules such that $\delta^2 \phi^A = 0$ as we can shift the
off-shell nilpotency problem of open algebras to the transformation rules
of the collective field. The originally present gauge symmetry can thus be
fixed in a manifestly BRST invariant way by adding $\delta \Psi(\phi)$. So,
the gauge fixed action can be decomposed as $S_{gf} = S_{inv} + \delta
\Psi$. The second requirement for a good quantisation scheme can be
satisfied by taking $S_{inv}$ to be BRST invariant, as Ward identities then
imply gauge independence.

Another restriction on $S_{inv}$ is that when used as weight in the
partition function (i.e. when putting $\Psi$ to zero), the original,
ill-defined path integrals are recovered. It is clear that
\beq
    S_{inv} = S_{BV} (\phi -\varphi ,\phi^*) - \phi^*_A c^A -\varphi^A B_A
\eeq
with the boundary condition that $S_{BV}(\phi, \phi^*=0) = S_0[\phi^i]$
satisfy this requirement. Moreover, imposing that the original, ill-defined
SD equations are recovered as Ward identities restrict us to this form
\cite{PoulSD}. We are now free to include in $S_{BV}$ whatever expansion in
the antifields $\phi^*$ we want, as they are fixed to zero anyway when
$\Psi=0$. The only condition is that $(S_{BV},S_{BV})=0$, as this
leads indeed to a BRST-invariant $S_{inv}$. The question whether open
algebras can be quantised in BV amounts then to proving that the classical
master equation can be solved for open algebras \cite{boekHenneaux}.

\section{Schwinger-Dyson Equations from two collective fields}

In this section, we will present the formalism with two collective fields
and derive the SD equation from them as a Ward identity without the
complication of gauge symmetries. In the derivation of the Ward identity,
we will already meet one peculiarity which will also be crucial in the next
section. We start from an action $S_0[\phi]$, depending on one bosonic
degree of freedom $\phi$. It has to be such that when exponentiated and put
under a path integral, it leads to a well-defined partition function and
perturbation series. We introduce two copies of the original field, the two
so-called collective fields, $\varphi_1$ and $\varphi_2$ and consider the
action $S_0[\phi -\varphi_1 -\varphi_2]$. This leads to two gauge
symmetries for which we introduce two ghostfields $\pi_1$ and $\phi^*_2$
and two antighost fields $\phi^*_1$ and $\pi_2$. The BRST--anti-BRST
transformation rules are
\beq
  \begin{array}{lcl}
     \delta_1 \phi = \pi_1 & \mbox{\hspace{2cm}} & \delta_2 \phi = \pi_2 \\
     \delta_1 \varphi_1 = \pi_1 - \phi^*_2 &  & \delta_2 \varphi_1 = -
\phi^*_1 \\
     \delta_1 \varphi_2 = \phi^*_2 & & \delta_2\varphi_2 = \pi_2 + \phi^*_1
\\
     \delta_1 \pi_1 = 0 &  & \delta_2 \pi_2 = 0 \\
     \delta_1 \phi^*_2 = 0 & & \delta_2 \phi^*_1 = 0.
\end{array}
\eeq

Imposing $(\delta_2 \delta_1 +\delta_1 \delta_2)\phi =0$ gives the  extra
condition $\delta_2 \pi_1 + \delta_1 \pi_2 =0$, while analogously
$(\delta_2 \delta_1 +\delta_1 \delta_2)\varphi_1 =0$ gives $\delta_1
\phi^*_1 +\delta_2 \phi^*_2 = \delta_2 \pi_1$.
$(\delta_2 \delta_1 +\delta_1 \delta_2)\varphi_2 =0$ leads to no new
condition. We introduce two extra bosonic fields $B$ and $\lambda$ and the
BRST transformation rules:
\beq
  \begin{array}{lcl}
      \delta_1 \pi_2 = B & \mbox{\hspace{2cm}} & \delta_2 \pi _1 = -B \\
      \delta_1 B = 0 &  & \delta_2 B = 0 \\
      \delta_1 \phi^*_1 = \lambda - \frac{B}{2} &  & \delta_2 \phi^*_2 =
-\lambda - \frac{B}{2} \\
      \delta_1 \lambda = 0 &  & \delta_2 \lambda = 0.
\end{array}
\eeq
All these rules together guarantee that $\delta_1^2 = \delta_2^2 = \delta
_1\delta_2 +\delta_2\delta_1 = 0$.

With all these BRST transformation rules at hand, we can construct a gauge
fixed action that is invariant under extended BRST symmetry. We will fix
both the collective fields to be zero. To that end, we add
\begin{eqnarray}
   S_{col} & = & \frac{1}{2} \delta_1 \delta_2 [ \varphi_1^2 - \varphi_2^2
] \nonumber \\
    & = & - (\varphi_1 + \varphi_2) \lambda+\frac{B}{2}(\varphi_1 - \varphi_2)
    + (-1)^a \phi^*_a \pi_a .
\end{eqnarray}
In the last term, there is a summation over $a=1,2$. Denoting
$\varphi_{\pm} = \varphi_1 \pm \varphi_2$, we have the gauge fixed action
\beq
   S_{gf} = S_0[\phi -\varphi_+]
 - \varphi_+\lambda+\frac{B}{2}\varphi_- + (-1)^a \phi^*_a \pi_a .
\eeq

The Schwinger-Dyson equations can be derived as Ward identities in the
following way.
\begin{eqnarray}
  0 & = & \langle \delta_1 [ \phi^*_1 F(\phi) ] \rangle \\
    & = & \int d\mu  \left[ \phi^*_1 \frac{\dr F}{\delta \phi} \pi_1 +
(\lambda - \frac{B}{2}) F(\phi) \right] e^{\ihbar S_{gf}}. \nonumber
\end{eqnarray}
We denoted the integration measure over all fields by $d\mu$. The term
$\langle B F(\phi) \rangle$ is zero. This can be seen by noticing that
$B = \delta_1\delta_2 \varphi_+$. The Ward identities themselves allow to
{\sl integrate by parts} to get
\beq
\langle B F(\phi) \rangle = - \langle \varphi_+ \delta_2 \delta_1 F(\phi)
\rangle,
\eeq
which drops out as $\varphi_+$ is fixed to zero. The same trick will be
usefull in deriving the Ward identities of the extended BRST symmetry in
the antifield scheme.

The SD equation then results as in \cite{PoulSD,PoulBV} by integrating out
$\pi_a$,$\phi^*_a$,$\lambda$,$B$,$\varphi_+$ and $\varphi_-$. Of course,
the SD equations can also be derived as Ward identities of the anti-BRST
transformation.

\section{Extended antifield formalism for closed, irreducible algebras}

The starting point is the same as in \cite{ik1}. Given any classical action
$S_0[\phi^i]$ with a closed and irreducible gauge algebra, the
configuration space is enlarged by introducing the
necessary ghosts, antighosts and auxiliary fields, as is described e.g. in
\cite{Baulieu}. The complete set of fields is
denoted by $\phi_A$ and their
BRST--anti-BRST transformation rules are all summarised by $\delta_a \phi_A
= \cR_{Aa} (\phi)$. For $a=1$, we have the BRST transformation rules, for
$a=2$ the anti-BRST transformation. Since the algebra is closed, we have
that
\beq
       \frac{\dr \cR_{Aa}(\phi)}{\delta \phi_B} \cR_{Ba}(\phi)  =  0
\eeq
and that
\beq
       \frac{\dr \cR_{A1}(\phi)}{\delta \phi_B} \cR_{B2}(\phi) +
       \frac{\dr \cR_{A2}(\phi)}{\delta \phi_B} \cR_{B1}(\phi)  =  0.
\eeq
In the first formula, there is no summation over $a$.

Instead of constructing a gauge fixed action that is invariant under the
extended BRST symmetry, we will introduce collective fields and associated
extra shift symmetries. Contrary to the previous collective field
approach to extended BRST invariant quantisation of \cite{ik1}, we
introduce {\it two} collective fields $\varphi_{A1}$ and $\varphi_{A2}$,
commonly denoted by $\varphi_{Aa}$, and replace everywhere $\phi_A$ by
$\phi_A - \varphi_{A1} - \varphi_{A2}$.
We now have two shift symmetries for which we
introduce the ghosts $\pi_{A1}$ and $\phi^{*2}_A$ with ghostnumber
$\gh{\pi_{A1}} = \gh{\phi^{*2}_A} = \gh{\phi_A} + 1$ and the antighosts
$\phi^{*1}_A$ and $\pi_{A2}$ with ghostnumber
$\gh{\pi_{A2}} = \gh{\phi^{*1}_A} = \gh{\phi_A} - 1$.
Again, we will use $\pi_{Aa}$ and $\phi^{*a}_A$ as compact notation.
Of course, one has to keep in mind that for $a=1$, $\pi_{Aa}$ is a ghost,
while for $a=2$, $\pi_{Aa}$ is an antighost and vice versa for $\phi^{*a}_A$.

We construct the BRST--anti-BRST transformations as follows:
\begin{eqnarray}
             \delta_a \phi_A & = & \pi_{Aa} \nonumber \\
      \delta_a \varphi_{Ab} & = & \delta_{ab} \left[ \pi_{Aa} - \epsilon
      _{ac} \phi^{*c}_{A} - \cR_{Aa}(\phi -\varphi_1 -\varphi_2) \right]
       + (1 - \delta_{ab}) \epsilon_{ac} \phi^{*c}_A ,
\end{eqnarray}
with no summation over
\footnote{Our convention: $\epsilon_{12} = 1$,$\epsilon^{12} = -1$.}
$a$ in the second line. These are chosen such that
\beq
      \delta_a(\phi_A - \varphi_{A1} -  \varphi_{A2}) =
      \cR_{Aa}(\phi -\varphi_1 -\varphi_2).
\eeq
The two collective fields lead to even more freedom than the one in the
collective field formalism for BV to shift the $\cR_{Aa}$ in the
transformation rules. However, we will see that the choice above leads to
what was already known as the antifield formalism for extended BRST
symmetry \cite{BLT}. Furthermore, the discussion of open algebras in
section 2 also indicates that it is useful to construct the rules such that
$\delta_a^2 \phi_A = 0$ and $(\delta_1 \delta_2 + \delta_2 \delta_1)\phi_A
= 0 $, independently of the closure of the algebra.
Imposing that $\delta_a^2 = 0$ ($a=1,2$) and that
$\delta _1\delta _2 +\delta _2 \delta_1 = 0 $ when acting on any field, we
are led to the introduction of two extra fields $B_A$ and $\lambda_A$ and
the new transformationrules:
\begin{eqnarray}
       \delta _a \pi_{Ab} & = & \epsilon_{ab} B_A \nonumber \\
       \delta_a B_A & = & 0 \\
       \delta_a \phi^{*b}_A &= & -\delta_a^b \left[ (-1)^a \lambda_A
        + \frac{1}{2} \left( B_A +
       \frac{\dr \cR_{A1}(\phi-\varphi_1 - \varphi_2)}{\delta \phi_B}
\cR_{B2}(\phi-\varphi_1 - \varphi_2 )  \right) \right] \nonumber \\
       \delta _a\lambda _A & = & 0  .\nonumber
\end{eqnarray}

Inspired by \cite{PoulBV,ik1}, we will gauge fix both the collective
fields to zero in a BRST--anti-BRST invariant way. For that purpose, we
need a matrix $M^{AB}$, with constant c-number entries and which is
invertible. Moreover, it has to have the symmetry property $M^{AB} =
(-1)^{\gras{A} \gras{B}} M_{BA}$ and all the entries of $M$ between
Grassmann odd and Grassmann even sectors have to vanish. It should be such
that $\phi_A M^{AB} \phi_B$ has over all ghostnumber zero and has even
Grassmann parity. Except for the constraints above, the precise form of $M$
is of no concern. It will drop out in the end completely \cite{ik1}.
The collective fields are then gauge fixed to zero by adding the term
\begin{eqnarray}
S_{col} & = & - \frac{1}{4} \epsilon^{ab} \delta_a \delta_b \left[
          \varphi_{A1} M^{AB} \varphi_{B1} - \varphi_{A2} M^{AB}
          \varphi_{B2} \right] \nonumber \\
       & = & - (\varphi_{A1} + \varphi_{A2}) M^{AB} \lambda_B
          + \frac{1}{2} ( \varphi_{A1} - \varphi_{A2}) M^{AB} B_B\\
      &  & + (-1)^{\gras{B}+1} \phi^{*1}_A M^{AB} \pi_{B1}
          + (-1)^{\gras{B}} \phi^{*2}_A M^{AB} \pi_{B2}  \nonumber \\
     &  & + (-1)^{\gras{B}} \phi^{*1}_A M^{AB} R_{B1} (\phi
-\varphi_1-\varphi_2) + (-1)^{\gras{B}+1} \phi^{*2}_A M^{AB} R_{B2} (\phi
-\varphi_1-\varphi_2)  \nonumber \\
     &  & +\frac{1}{2} (\varphi_{A1} - \varphi_{A2}) M^{AB}
       \frac{\dr \cR_{B1}(\phi-\varphi_1 - \varphi_2)}{\delta \phi_C}
\cR_{C2}(\phi-\varphi_1 - \varphi_2 ) \nonumber.
\end{eqnarray}
The relative sign between the two contributions of the gauge fixing is
needed to make two terms containing the product $\phi^{*1}_A M^{AB}
\phi^{*2}_B$, cancel. Redefine now $\varphi_{A\pm} = \varphi_{A1}
\pm \varphi_{A2}$, which allows us to rewrite the gauge fixing terms
in a more compact and suggestive form:
\begin{eqnarray}
S_{col} & = & - \varphi_{A+} M^{AB} \lambda_B + \frac{1}{2} \varphi_{A-}
         M^{AB} B_B + (-1)^a (-1)^{\gras{B}} \phi^{*a}_{A} M^{AB} \pi_{Ba}
         \nonumber \\
     &  & +\frac{1}{2} \varphi_{A-}  M^{AB}
       \frac{\dr \cR_{B1}(\phi-\varphi_+)}{\delta \phi_C}
\cR_{C2}(\phi-\varphi_+)  \\
    &   & + (-1)^{a+1}
      (-1)^{\gras{B}} \phi^{*a}_A M^{AB} R_{Ba} (\phi -\varphi_+).
      \nonumber
\end{eqnarray}
Notice that this time a summation over $a$ {\it is} understood in the third
and fifth term. The $\phi^{*a}_A$ have indeed become source terms for the
BRST and anti-BRST transformation rules, while the difference of the two
collective fields $\varphi_{A-}$ acts as a source for mixed
transformations. The sum of the two collective fields is just fixed to
zero.

The original gauge symmetry can be fixed in an extended BRST invariant way
by adding the variation of a gauge boson $\Psi(\phi)$, of ghostnumber zero.
We take it to be only a function of the original fields $\phi_A$. This
gives the extra terms
\begin{eqnarray}
S_{\Psi} & = & \frac{1}{2} \epsilon^{ab} \delta_a \delta_b \Psi(\phi)
           \nonumber \\
         & = & - \frac{\dr \Psi}{\delta \phi_A} B_A + \frac{1}{2}
\epsilon^{ab} (-1)^{\gras{B}+1} \left[ \frac{\dr}{\delta \phi_A}
\frac{\dr}{\delta \phi _B} \Psi \right] . \pi_{Aa} \pi_{Bb}.
\end{eqnarray}

In order to show that we now have the antifield formalism which was derived
on algebraic grounds in \cite{BLT}, we first have to make the following
(re)definitions.
We incorporate the matrix $M^{AB}$ introduced above in the antifields:
\begin{eqnarray}
\phi ^{*Aa'} &=& (-1)^{\gras{A}}\phi ^{*\ a}_B
M^{BA}(-1)^{a+1}\hspace*{1cm}a = 1,2\nonumber\\
\mbox{}\\
\bar \phi ^A &=& \frac{1}{2} \varphi _{B-} M^{BA} . \nonumber
\label{redef}
\end{eqnarray}
Owing to the properties of the matrix $M^{AB}$ above, the ghostnumber
assignments after the redefinition are given by
\begin{eqnarray}
\gh{\phi ^{*Aa'}} &=& (-1)^a - \gh{\phi _A} \nonumber \\
\gh{\bar \phi ^A} &=& -\gh{\phi _A},
\end{eqnarray}
while the Grassmann parities are of course
\beq
\varepsilon _{\phi ^{*Aa'}} = \varepsilon _{\phi _A} + 1\ \ ;\ \
\varepsilon _{\bar \phi ^A} = \varepsilon _{\phi _A}\,.
\eeq
We denote the so-called {\it extended action} of Batalin, Lavrov and Tyutin
\cite{BLT} by $S_{BLT}$.  Using the new variables and dropping the primes,
it is defined by
\beq
S_{BLT}(\phi _A,\phi ^{*Aa},\bar \phi ^A) = S_0[\phi _A]
+ \phi ^{*Aa}R_{Aa}(\phi ) + \bar \phi ^A \frac{\dr
 R_{A1}(\phi )}{\delta \phi _B}R_{B2}(\phi )\,.
\label{SBLT}
\eeq
The remaining terms of $S_{col}$, we denote by $S_\delta $, hence
\beq
S_\delta = -\varphi _{A+}M^{AB}\lambda _B
   + \bar \phi ^AB_A - \phi ^{*Aa}\pi _{Aa}\,.
\eeq
The notation stems from the fact that integrating over $\pi _{Aa}$, $B_A$
and $\lambda _B$ leads to a set of $\delta $-functions removing all the
terms originating in the collective field formalism.  The situation is then
analogous to the BV scheme.  Before the gauge fixing term $S_\Psi $ is
added, all antifields are fixed to zero.

With all these definitions at hand, we have that
\begin{eqnarray}
\label{Sgf}
S_{gf} &=& S_0[\phi - \varphi _+] + S_{col} + S_\Psi\\
&=& S_{BLT}[\phi - \varphi _+,\phi ^{*a},\bar \phi ] + S_\delta + S_\Psi
\,,\nonumber
\end{eqnarray}
which gives the gauge fixed partition function
\beq
\cZ = \int [d\phi ][d\phi ^{*a}][d\bar \phi ][d\pi a][dB]
 e^{\frac{i}{\hbar }S_{BLT}[\phi ,\phi ^{*a},\bar \phi
]}e^{\frac{i}{\hbar }S_\Psi }e^{\frac{i}{\hbar }\tilde S_\delta }\,.
\eeq
We already integrated out $\lambda $ and $\varphi _+$.  Hence, $\tilde
S_\delta $ is $S_\delta $ with the $-\varphi _{A+}M^{AB}\lambda _B$
omitted.  The gauge fixing term $e^{\frac{i}{\hbar }S_\Psi }$ can be
obtained by acting with an operator $\hat V$ on $e^{\frac{i}{\hbar
}\tilde S_\delta }$, i.e.
\beq
e^{\frac{i}{\hbar }S_\Psi }e^{\frac{i}{\hbar }\tilde S_\delta } = \hat
V e^{\frac{i}{\hbar }\tilde S_\delta }\,.
\eeq
{}From the explicit form of $\tilde S_\delta $ and $S_\Psi $, and using
that $e^{a(y)\frac{\delta }{\delta x}}f(x) = f(x + a(y))$,
we see that
$ \hat V(\Psi) = e^{-T_1(\Psi )- T_2(\Psi)}$ with
\begin{eqnarray}
T_1 (\Psi ) & = & \frac{\dr \Psi (\phi )}{\delta \phi
_A}\,\cdot\,\frac{\vec \delta }{\delta \bar \phi ^A} \nonumber \\
T_2(\Psi)  & = & \frac{i\hbar
}{2}\varepsilon ^{ab}\frac{\vec\delta }{\delta \phi ^{*Bb}}\frac{\dr
 }{\delta \phi_A}\frac{\dr}{\delta \phi_B}\Psi
\frac{\vec \delta }{\delta \phi ^{*Aa}}\,.
\end{eqnarray}
The convention is that derivatives act on everything standing on the right
of them.  The operator $\hat V$ can be integrated by parts, such that
\beq
\cZ = \int [d\phi ][d\phi ^{*a}][d\bar \phi ]\left[\hat U(\Psi )
e^{\frac{i}{\hbar }S_{BLT}}\right] \delta (\phi ^{*A1}) \delta (\phi
^{*A2}) \delta (\bar \phi ^A)\, ,
\eeq
with the operator $\hat U = e^{ + T_1 - T_2 }$.
This form of the path integral agrees with \cite{BLT}.
The quantisation prescription is then to construct $S_{BLT}$,
function of fields and antifields.  Then the gauge fixing is done by acting
with the operator $\hat U(\Psi )$.  Then the antifields $\phi ^{*Aa}$ and
$\bar \phi ^A$ are removed by the $\phi $-functions which fix them to zero.
This is the most important difference with the collective field formalism
for extended BRST symmetry in \cite{ik1}.  There the antifields $\phi
^{*Aa}$ were removed by a Gaussian integral, and it is precisely this
procedure which prevented the generalisation of the results of \cite{ik1}
to open algebras \cite{ik2}. Notice however that instead of acting with
$\hat U$ on $e^{\ihbar S_{BLT}}$, it is a lot easier to take as realisation
of the gauge fixing $S_{\Psi} + \tilde S_{\delta}$, especially when
$S_{BLT}$ becomes non-linear in the antifields.

Let us finally derive the {\it classical master equations} which are
satisfied by $S_{BLT}$.  They follow from the fact that $S_{gf}$
(\ref{Sgf}) is invariant under both the BRST and the anti-BRST
transformation.  Furthermore, one has to use the fact that the matrix
$M^{AB}$ only has non-zero entries for $\varepsilon _A = \varepsilon _B$,
and hence that $M^{AB} = (-1)^{\varepsilon_ A} M^{BA} = (-1)^{\varepsilon_
B} M^{BA}$.  Also, in the collective field BRST transformation rules, we
may replace $R_{Aa}(\phi - \varphi _+)$ by
$\frac{\vec\delta S_{BLT}}{\delta \phi ^{*Aa'}}$.
Using that $\delta _aS_\Psi = 0$ on itself, we have that
\begin{eqnarray}
0 &=& \delta _a S_{gf}  \nonumber \\
&=& \delta _a S_{BLT} + \delta _a S_\delta \\
&=& \frac{\dr S_{BLT}}{\delta \phi _A}\,\cdot\,\frac{\vec\delta
S_{BLT}}{\delta \phi ^{*Aa'}} + \varepsilon _{ab} \phi ^{*Ab}
\frac{\vec\delta S_{BLT}}{\delta \bar \phi ^A} \nonumber
\end{eqnarray}
We introduce two antibrackets, one for every $\phi ^{*Aa}$, defined by
\beq
(F,G)_a = \frac{\dr F}{\delta \phi
_A}\,\cdot\,\frac{\vec\delta G}{\delta \phi ^{*Aa}} - \frac{\dr
F}{\delta \phi ^{*Aa}}\,\cdot\,\frac{\vec\delta G}{\delta \phi _A}\,.
\eeq
Of course, they have the same properties as the antibrackets from the usual
BV scheme, so that we finally can write the classical master equations as
\beq
\frac{1}{2}(S_{BLT},S_{BLT})_a + \varepsilon _{ab} \phi ^{*Ab}
\frac{\vec\delta S_{BLT}}{\delta \bar \phi ^A} = 0\,.
\label{ClasBLT}
\eeq
For closed, irreducible algebras, we know that the solution is of the form
(\ref{SBLT}).

\section{Ward's Identities and Quantum Master Equation}

In this section, we first derive the Ward identities for the extended BRST
symmetry and then we take these identities as a starting point to derive
the quantum master equation.

\subsection{Ward's Identities}

As the gauge fixed action we constructed (\ref{Sgf}) is invariant under
both the BRST and anti-BRST transformation rules, the standard procedure
based on Shakespeare's theorem \cite{ik3} allows the derivation of 2 types
of Ward identities.  For any $X$, we have that
\begin{eqnarray}
\langle \delta _1 X \rangle &=& 0 \nonumber \\
\langle \delta _2 X \rangle &=& 0\,,
\end{eqnarray}
where $\langle {\cal A} \rangle$ means the quantum expectation value using
the gauge fixed action (\ref{Sgf}) of an operator $\cA$.  As we are only
interested in the
theory after having integrated out $\varphi_+$, we will restrict ourselves
to quantities $X(\phi _A,\phi ^{*Aa},\bar \phi )$.  Furthermore, we assume
that the quantum corrections~- the counterterms~- which may be needed to
cancel the non-invariance of the measure and hence to guarantee the
validity of the Ward identities, do not spoil the gauge fixing procedure
described above.  Like in the case of BV in section 2, if they would, the
derivation of the SD equations as Ward identities would be invalidated.
Hence, they are restricted to be of the form $M(\phi -
\varphi _+,\phi ^{*Aa},\bar \phi )$ and
\beq
W_{BLT}(\phi ,\phi ^{*Aa},\bar \phi ) = S_{BLT}(\phi ,\phi ^{*Aa},\bar
\phi ) + \hbar M(\phi ,\phi ^{*Aa},\bar \phi )\,.
\eeq
The Ward identities then become
\begin{eqnarray}
0 &=& \langle \delta a X \rangle\\
&=& \int [d\phi ][d\phi ^{*a}][d\bar \phi ][d\varphi _+][d\pi
_a][dB][d\lambda ] \delta _a X\,\cdot\,e^{\frac{i}{\hbar
}W_{BLT}(\phi - \varphi _+,\phi ^{*Aa},\bar \phi )}\\
& & \hspace*{1cm}\cdot\,\hat V \left[e^{\frac{i}{\hbar }\tilde
S_\delta
}\right]\,\cdot\,e^{-\frac{i}{\hbar }\varphi _{A+}M^{AB}\lambda _B}\,.
\nonumber
\end{eqnarray}
Let us take $a = 1$.  Then
\begin{eqnarray}
\delta _1 X &=& \frac{\dr X}{\delta \phi _A}\,\cdot\,\pi _{A1} +
\frac{\dr X}{\delta \phi ^{*A1'}}(-1)^{\gras{A}} M^{BA} \left[\lambda _B -
\frac{1}{2}\left(B_B + \frac{\dr R_{B1}}{\delta \phi
_C}R_{C2}\right)\right] \nonumber \\
 & & + \frac{\dr X}{\delta \bar \phi _A}\,\cdot\,\frac{1}{2}
M^{BA} [-2\phi _B^{*2} + \pi _{B1} - R_{B1}(\phi - \varphi _+)]\,.
\end{eqnarray}
We reintroduced the primes for the $\phi ^{*Aa'}$ in order to distinguish
the antifields before and after the redefinition.  Now it is important to
notice that
\begin{eqnarray}
\delta _a \varphi _{A+} &=& \pi_{Aa} - R_{Aa}(\phi - \varphi _+)\\
- \delta _2 \delta _1 \varphi _{A+} &=& -\delta _2 (\pi _{A1} - R_{A1}(\phi
- \varphi _+)) = B_A + \frac{\dr  R_{A1}(\phi - \varphi _+)}{\delta
\phi _B } R_{B2}(\phi  - \varphi _+)\,.
\nonumber
\end{eqnarray}
Using this, and noticing that the Ward identities themselves allow us to
`integrate by parts' the (anti-)BRST operator, we see that
\begin{eqnarray}
\left\langle \frac{\dr X}{\delta \phi ^{*A1'}} \left(B_B +
\frac{\dr R_{B1}}{\delta \phi _C} R_{C2}\right)\right \rangle &=&
\left\langle \delta _1\delta _2 \left[\frac{\dr X}{\delta \phi
^{*A1}}\right] \varphi _{B+}\right\rangle\\
\left\langle \frac{\dr X}{\delta \bar \phi _A} (\pi _{B1} -
R_{B1})\right\rangle &=& (-1)^{\gras{B}+1} \left\langle \delta _1
\left[\frac{\dr X}{\delta \bar \phi _A}\right] \varphi
_{B+}\right\rangle\,.
\nonumber
\end{eqnarray}
Hence, both terms disappear from the Ward identity as $\varphi _+$ is fixed
to zero by a delta function integration.  Denote the complete measure of
the path integral by $d\mu $, then we can write the remaining Ward identity
as
\begin{eqnarray}
0 &=& \int d\mu \left[\frac{\dr X}{\delta \phi _A} \pi _{A1} +
\frac{\dr X}{\delta \phi ^{*A1'}} (-1)^{\gras{A}} M^{BA} \lambda _B + \phi
^{*A2'} (-1)^{\gras{X}} \frac{\vec \delta X}{\delta \bar \phi _A}\right]
\nonumber \\
& & \cdot e^{\frac{i}{\hbar }W_{BLT}(\phi - \varphi _+,\phi ^{*Aa},\bar
\phi )} \hat V \left[e^{\frac{i}{\hbar }\tilde S_\delta }\right]
e^{-\frac{i}{\hbar }\varphi _{A+} M^{AB} \lambda _B}\,.
\end{eqnarray}
In the first term, the $\varphi _{A+}$ can trivially be integrated out.
Then, considering the expressions for $\hat V$ and $\tilde S_\delta $, we
see that $\pi _{A1}$ can be replaced by $-\frac{\hbar
}{i}\,\frac{\vec\delta }{\delta \phi ^{*A1}}$.
We then integrate by parts over $\phi ^{*A1}$,
which leads to
\begin{eqnarray}
&&\frac{\hbar }{i} (-1)^{\varepsilon _X(\varepsilon _A + 1)}
\frac{\vec\delta }{\delta \phi ^{*A1}} \left[\frac{\dr X}{\delta
\phi _A} e^{\frac{i}{\hbar }W_{BLT}}\right] \hat V
\left[e^{\frac{i}{\hbar }\tilde S_\delta }\right]\\
&&= \left[-i\hbar \Delta _1 X + \frac{\dr X}{\delta \phi
_A}\,\cdot\,\frac{\vec\delta W_{BLT}}{\delta \phi ^{*A1}}\right]
 e^{\frac{i}{\hbar }W_{BLT}} \hat V \left[e^{\frac{i}{\hbar }\tilde
S_\delta }\right]
\nonumber
\end{eqnarray}
under the path integral.  Here, we generalised that other operator
well-known from BV:
\beq
\Delta_a X = (-1)^{\varepsilon _A + 1} \frac{\dr }{\delta \phi
^{*Aa}}\ \frac{\dr}{\delta \phi _A} X\,.
\eeq
For the second term we can proceed analogously by replacing $M^{AB} \lambda
_B e^{- \frac{i}{\hbar }\phi _{A+}M^{AB}\lambda _B}$ by
$\left(-\frac{\hbar }{i}\right) \frac{\vec\delta }{\delta \varphi
_{A+}}e^{-\frac{i}{\hbar }\varphi _{A+}M^{AB}\lambda _B}$.
Integrating by parts over $\varphi _{A+}$, we see that the derivative can
only act on $W_{BLT}(\phi - \varphi _+,\phi ^{*a},\bar \phi )$, and we get
under the path integral
\beq
\frac{\dr X}{\delta \phi ^{*A1'}}\ \frac{\hbar }{i}\
\frac{\vec\delta }{\delta \varphi _{A+}}\ e^{\frac{i}{\hbar } W_{BLT} (\phi
- \varphi _+,\phi ^{*a},\bar \phi )} \hat V \left[e^{\frac{i}{\hbar
}\tilde S_\delta }\right]\delta (\varphi _+)\,.
\eeq
Remembering that $\frac{d}{dx} f(x - y) = -\frac{d}{dy} f(x - y)$, this
leads finally to
\beq
\int d\mu - \frac{\dr X}{\delta \phi ^{*A1'}}\
\frac{\vec\delta W_{BLT}}{\delta \phi _A}\  \hat
V \left[e^{\frac{i}{\hbar }\tilde S_\delta} \right]\,.
\eeq
The complete Ward identity hence becomes, dropping the primes,
\begin{eqnarray}
0 &=& \left\langle (X,W_{BLT})_1 - i\hbar \Delta _1 X + (-1)^{\varepsilon
_X} \phi ^{*A2} \frac{\vec\delta X}{\delta \bar \phi _A}\right\rangle\\
&=& \int [d\phi ][d\phi ^{*a}][d\bar \phi ]\left[(X,W_{BLT})_1 -
i\hbar \Delta _1 X + (-1)^{\varepsilon
_X} \phi ^{*A2} \frac{\vec\delta X}{\delta \bar \phi _A}\right]
\nonumber
\end{eqnarray}
\[e^{\frac{i}{\hbar }W_{BLT}} \hat V \left[e^{\frac{i}{\hbar }\tilde
S_\delta }\right]\,.\]
An analogous property is of course obtained by going through the same steps
for the Ward identities $\langle \delta _2 X \rangle = 0$.

\subsection{Quantum Master Equation}

As in the case of the BV formalism, the fact that this Ward identity is
valid for all $X(\phi ,\phi ^{*a},\bar \phi )$ leads to an equation on
$W_{BLT}$, the so-called {\it quantum master equation}. Starting from the
most general Ward identity, the purpose is of course to remove all
derivative operators acting on $X$ by partial integrations. Again, we
denote by $d\mu$ the measure of the path integral. We thus start from
\begin{eqnarray}
   0 & = & \int d\mu \left[ \frac{\dr X}{\delta \phi_A} \frac{\dl
W_{BLT}}{\delta \phi^{*Aa}} - \frac{\dr X}{\delta \phi^{*Aa}}\frac{\dl
W_{BLT}}
{\delta \phi_A} -i\hbar (-1)^{\gras{A}+1} \frac{\dr}{\delta \phi^{*Aa}}
\frac{\dr}{\delta \phi_A} X \right. \nonumber \\
 & & \left. + (-1)^{\gras{X}} \epsilon_{ab} \phi^{*Ab} \frac{\dl X}{\delta
\bar \phi _A} \right] e^{\ihbar(W_{BLT} + S_{\Psi} + \tilde S_{\delta})}.
   \label{Ward}
\end{eqnarray}
Notice that we have reexpressed the operator $\hat V$ as
$e^{\ihbar S_{\Psi}}$.

By integrating by parts over $\phi_A$ in the first term, we get the
following two terms:
\begin{eqnarray}
 & &   \int d\mu \, i\hbar X \Delta_a e^{\ihbar W_{BLT}} . e^{\ihbar(
S_{\Psi} + \tilde S_{\delta})} \nonumber \\
 &  & + \int d\mu \, i\hbar X (-1)^{\gras{A}+1} \frac{\dr e^{\ihbar
W_{BLT}}} {\delta \phi^{*Aa}} \frac{\dr}{\delta \phi_A} \left[
 e^{\ihbar( S_{\Psi} + \tilde S_{\delta})} \right].
 \label{Term1}
\end{eqnarray}
The second and third contribution to the Ward identity (\ref{Ward}) can be
combined to give
\beq
   \int d\mu  (-i\hbar) (-1)^{\gras{A}+1} \frac{\dr}{\delta \phi_A} \left[
   \frac{\dr X}{\delta \phi^{*Aa}} e^{\ihbar W_{BLT}} \right]
 e^{\ihbar( S_{\Psi} + \tilde S_{\delta})} .
\eeq
Integrating by parts twice, first over $\phi_A$, then over $\phi^{*Aa}$
gives us the terms:
\begin{eqnarray}
 &  & \int d\mu \, i\hbar (-1)^{\gras{A}} X e^{\ihbar W_{BLT}}
\frac{\dr}{\delta \phi^{*Aa}} \frac{\dr}{\delta \phi_A} \left[
 e^{\ihbar( S_{\Psi} + \tilde S_{\delta})} \right]  \nonumber \\
 & + & \int d\mu \, i\hbar (-1)^{\gras{A}} X \frac{\dr e^{\ihbar
W_{BLT}}}{\delta \phi^{*Aa}} . \frac{\dr}{\delta \phi_A}
\left[ e^{\ihbar( S_{\Psi} + \tilde S_{\delta})} \right].
\label{Term2}
\end{eqnarray}
Notice that the second term of (\ref{Term1}) cancels the second term of
(\ref{Term2}).

Also in the fourth term of (\ref{Ward}), we have to integrate by parts,
over $\bar \phi^A$. This gives us again two terms:
\begin{eqnarray}
 & - &\int  d\mu \, X \epsilon_{ab} \phi^{*Ab}
  \frac{\dl e^{\ihbar W_{BLT}}}{\delta \bar \phi^A}
   e^{\ihbar( S_{\Psi} + \tilde S_{\delta})} \nonumber \\
 & - & \int d\mu \, X \epsilon_{ab} \phi^{*Ab} e^{\ihbar W_{BLT}} \frac{\dl
 }{\delta \bar\phi^A}
\left[ e^{\ihbar( S_{\Psi} + \tilde S_{\delta})} \right].
\label{Term3}
\end{eqnarray}
It is possible to show that the first term in (\ref{Term2}) and the
second term in (\ref{Term3}) cancel. Working out the two derivatives and
using the explicit form of $\tilde S_{\delta}$, we rewrite the first term
of (\ref{Term2}) as
\beq
   \int d\mu  (i\hbar) (\ihbar)^2 X e^{\ihbar W_{BLT}}
 e^{\ihbar( S_{\Psi} + \tilde S_{\delta})}
 \frac{\dr S_{\Psi}}{\delta \phi_A} \pi_{Aa}.
\eeq
Now, we know that $\delta_a S_{\Psi} = 0$, which allows us to replace
$ \frac{\dr S_{\Psi}}{\delta \phi_A} \pi_{Aa}$ by $- \frac{\dr
S_{\Psi}}{\delta \pi_{Ab}} \epsilon_{ab} B_A$. Using the explicit form of
$ \tilde S_{\delta}$ again, this is
\beq
   - \int d\mu  (i\hbar) X e^{\ihbar W_{BLT}} \frac{\dr e^{\ihbar
S_{\Psi}}}{\delta \pi_{Ab}} .\frac{\dl e^{\ihbar \tilde S_{\delta}}}{\delta
\bar \phi^A} \epsilon_{ab}.
\eeq
One more partial integration, over $\pi_{Ab}$, is needed to see that the
terms do cancel as mentioned above.

Summing all this up, we see that the Ward identities are equivalent to
\beq
  0 = \int d\mu  X \left[ i\hbar \Delta_a - \epsilon_{ab} \phi^{*Ab}
\frac{\dl}{\delta \bar\phi^A} \right] e^{\ihbar W_{BLT}} \hat V
e^{\ihbar \tilde S_{\delta}}.
\eeq
As this is valid for all possible choices for $X(\phi,\phi^{*a},\bar\phi)$,
we see that $W_{BLT}$ has to satisfy the {\it quantum master equation}
\beq
   \left[ i\hbar \Delta_a - \epsilon_{ab} \phi^{*Ab}
\frac{\dl}{\delta \bar\phi^A} \right] e^{\ihbar W_{BLT}} = 0.
\eeq
This is equivalent to
\beq
   \frac{1}{2} (W_{BLT},W_{BLT})_a + \epsilon_{ab} \phi^{*Ab} \frac{\dl
W_{BLT}}{\delta \bar \phi^A} = i\hbar \Delta_a W_{BLT}.
\eeq
Remember that these are two equations, $a=1,2$. By doing the usual
expansion $W_{BLT} = S_{BLT} + \hbar M_1 + \hbar^2 M_2 + \ldots$, we
recover the classical master equation (\ref{ClasBLT}) for $S_{BLT}$.

\section{Open Algebras}

In section 2, we pointed out how combining the collective field approach
and the recipe of \cite{Open}, one is naturally lead to the construction of
an extended action that contains terms of quadratic and higher order in the
antifields. As we do not have a principle analogous to \cite{Open} for
constructing a gauge fixed action that is invariant under extended BRST
symmetry for the case of an open algebra, we will have to take the other
point of view advocated there.

Although we may compare the collective field method
{\sl to a method sometimes employed in French
cuisine : a piece of pheasant meat is cooked between two slices of veal,
which are then discarded} \cite{GM}, the collective fields again
play an important part.
Like in the case of ordinary BRST collective field quantisation, the
introduction of the collective fields allow
us to shift the problem of the off-shell non-nilpotency to the (anti-)BRST
transformations of the collective fields. Indeed, $\delta_a \phi_A =
\pi_{Aa}$,
$\delta_a \pi_{Ab} = \epsilon_{ab} B_A$ and $\delta_a B_A = 0$ guarantee
that $\delta_a^2 \phi_A = 0 $  and that $(\delta_1 \delta_2 + \delta_2
\delta_1)\phi_A = 0$. Therefore, the originally present gauge symmery can
be fixed in an extended BRST invariant way
like for closed algebras, i.e. by adding $S_{\Psi} = \frac{1}{2}
\epsilon^{ab} \delta_a \delta_b \Psi$ to an extended BRST invariant action,
$S_{inv}$. This way, Ward's identities guarantee that whatever way we
choose to construct $S_{inv}$, the partition function will be invariant of
the gauge choice if $S_{inv}$ is extended BRST invariant.

Another requirement that has to be satisfied by a good quantisation
procedure, is that when the gauge fixing is omitted, one gets the original,
ill-defined partition function back. It is clear that by decomposing
$S_{inv} = S_{BLT} + \tilde S_{\delta}$, with the familiar form for $\tilde
S_{\delta}$ and with $S_{BLT} = S_0 + \ldots $ where the dots stand for
terms of at least first order in the antifields $\phi^{*Aa}$ and $\bar
\phi^A$, does satisfy that requirement. Imposing that the SD equations
are derivable as Ward identities again restricts us to this form. The naive
point of view is then
that before we add the gauge fixing $S_{\Psi}$, the antifields are fixed to
zero by $\tilde S_{\delta}$, and we can hence add whatever terms
proportional to them.

As far as the invariance of $S_{inv}$ under extended BRST transformations
is concerned, we know that that is indeed satisfied if we take $S_{BLT}$ to
be a solution of the classical master equation (\ref{ClasBLT}) and take
\begin{eqnarray}
 \delta_a\varphi_{Ab} & = & \delta_{ab} \left[ \pi_{Aa} - \epsilon_{ac}
\phi^{*c}_A - \frac{\dl S_{BLT}(\phi - \varphi_+)}{\delta \phi^{*Aa}}
\right] + (1 - \delta_{ab}) \epsilon_{ac} \phi^{*c}_A \nonumber \\
  \delta_a \phi^{*b}_A & = & - \delta_a^b \left[ (-1)^a \lambda_A +
\frac{1}{2} (B_A + \frac{\dl S_{BLT}(\phi -\varphi_+)}{\delta \bar
\phi^A})\right].
\end{eqnarray}
Hence, we see that the question whether open algebras can be quantised in
an extended BRST invariant way, reduces to the fact whether a solution to
(\ref{ClasBLT}) can be found for open algebras with the boundary
condition
that $S_{BLT} = S_0 + \phi^{*Aa} \cR_{Aa} + \ldots$ It has been proved that
such solutions exist \cite{BLT,Gregoire,Brussel}.

As far as the treatment of reducible gauge algebras is concerned, the
collective field formalism does not define the ghostspectrum that has to be
introduced for a correct quantisation. As was pointed out already in
\cite{ik2}, once the configuration space is constructed correctly for a
reducible gauge algebra, one is left with a nilpotent or on-shell nilpotent
set of extended BRST transformation rules. Both cases are in fact the ones
treated above.

\section{Conclusion}

In this paper, we modified the collective field approach to quantisation of
gauge theories in order to derive an antifield formalism for extended BRST
invariant quantisation. We introduced two collective fields for every field.
This way, we have two ghost-antighost pairs associated with the two shift
symmetries. The antighost field of the first pair acts as a source for the
BRST transformations, the ghost field of the second as a source for the
anti-BRST transformations. The remaining ghost and antighost naturally lead
to a representation of the gauge fixing. The sum of the two collective
fields can be integrated out trivially, while their difference is needed as
a sourceterm for the composition of a BRST and and anti-BRST
transformation. We argue that this approach does allow for the extended
BRST invariant treatment of open algebras, stressing the importance of the
part played by the collective fields.

\section*{Acknowledgment}
I sincerely thank P.H. Damgaard for very useful comments on an early draft.

\end{document}